\documentstyle[11pt,aaspp4]{article}

\newcommand{\noun}[1]{\textsc{#1}}

\begin{document}

\newcommand{\La}{\mathrm{Ly}\alpha }

\newcommand{\Lalpha}{\mathrm{Ly}\alpha \, \lambda 1215}

\newcommand{\Ka}{\mathrm{K}\alpha }

\newcommand{\Lb}{\mathrm{L}\beta }

\newcommand{\Ha}{\mathrm{H}\alpha }

\newcommand{\Halpha}{\mathrm{H}\alpha \, \lambda 6563}

\newcommand{\Hb}{\mathrm{H}\beta }

\newcommand{\Hbeta}{\mathrm{H}\beta \, \lambda 4861}

\newcommand{\Pa}{\mathrm{P}\alpha }

\newcommand{\HeI}{\mathrm{He}\, {\textsc {i}}\, \lambda 5876}

\newcommand{\HeII}{\mathrm{He}\, {\textsc {ii}}\, \lambda 1640}

\newcommand{\CIII}{\mathrm{C}\, {\textsc {iii}}\, \lambda 977}

\newcommand{\CIIIb}{\mathrm{C}\, {\textsc {iii]}}\, \lambda 1909}

\newcommand{\CIV}{\mathrm{C}\, {\textsc {iv}}\, \lambda 1549}

\newcommand{\bOIIIbA}{[\mathrm{O}\, {\textsc {iii]}}\, \lambda 4363}

\newcommand{\bOIIIbB}{[\mathrm{O}\, \textsc {iii]}\, \lambda 5007}

\newcommand{\OIIIb}{\mathrm{O}\, {\textsc {iii]}}\, \lambda 1663}

\newcommand{\OVb}{\mathrm{O}\, {\textsc {v]}}\, \lambda 1218}

\newcommand{\OVI}{\mathrm{O}\, {\textsc {vi}}\, \lambda 1035}

\newcommand{\OIVb}{\mathrm{O}\, {\textsc {iv]}}\, \lambda 1402}

\newcommand{\bOIIb}{[\mathrm{O}\, {\textsc {ii]}}\, \lambda 3727}

\newcommand{\bOIb}{[\mathrm{O}\, {\textsc {i]}}\, \lambda 6300}

\newcommand{\NV}{\mathrm{N}\, {\textsc {v}}\, \lambda 1240}

\newcommand{\NIVb}{\mathrm{N}\, {\textsc {iv]}}\, \lambda 1486}

\newcommand{\NIIIb}{\mathrm{N}\, {\textsc {iii]}}\, \lambda 1750}

\newcommand{\MgII}{\mathrm{Mg}\, {\textsc {ii}}\, \lambda 2798}

\newcommand{\bNeVb}{[\mathrm{Ne}\, {\textsc {v]}}\, \lambda 3426}

\newcommand{\NeVIII}{\mathrm{Ne}\, {\textsc {viii}}\, \lambda 774}

\newcommand{\SiIV}{\mathrm{Si}\, {\textsc {iv}}\, \lambda 1397}

\newcommand{\bFeXb}{[\mathrm{Fe}\, {\textsc {x]}}\, \lambda 6734}

\newcommand{\bFeXIb}{[\mathrm{Fe}\, {\textsc {xi]}}\, \lambda 7892}

\newcommand{\FeII}{\mathrm{Fe}\, {\textsc {ii}}\, }

\newcommand{\bSiVIIb}{[\mathrm{Si}\, {\textsc {vii]}\, \lambda 2.5\mu \mathrm{m}}}

\newcommand{\bSiIXbA}{[\mathrm{Si}\, {\textsc {ix]}}\, \lambda 2.6\mu \mathrm{m}}

\newcommand{\bSiIXbB}{[\mathrm{Si}\, {\textsc {ix]}}\, \lambda 3.9\mu \mathrm{m}}

\newcommand{\bMgVIIIb}{[\mathrm{Mg}\, {\textsc {viii]}}\, \lambda 3.0\mu \mathrm{m}}

\newcommand{\bMgIVb}{[\mathrm{Mg}\, {\textsc {iv]}}\, \lambda 4.5\mu \mathrm{m}}
 
\newcommand{\bMgVIIb}{[\mathrm{Mg}\, {\textsc {vii]}\, \lambda 5.5\mu \mathrm{m}}}

\newcommand{\bNeVIb}{[\mathrm{Ne}\, {\textsc {vi]}}\, \lambda 7.6\mu \mathrm{m}}

\newcommand{\bArIIIb}{}
 
\newcommand{\bArVIb}{[\mathrm{Ar}\, {\textsc {vi]}}\, \lambda 4.5\mu \mathrm{m}}
 
\newcommand{\bSIVb}{[\mathrm{S}\, {\textsc {iv]}}\, \lambda 10.5\mu \mathrm{m}}

\newcommand{\bNeIIb}{[\mathrm{Ne}\, {\textsc {ii]}}\, \lambda 12.8\mu \mathrm{m}}

\newcommand{\bNeVbA}{[\mathrm{Ne}\, {\textsc {vi]}}\, \lambda 14.3\mu \mathrm{m}}

\newcommand{\bNeIIIbA}{[\mathrm{Ne}\, {\textsc {iii]}}\, \lambda 15.6\mu \mathrm{m}}

\newcommand{\bSIIIbA}{[\mathrm{S}\, {\textsc {iii]}}\, \lambda 18.7\mu \mathrm{m}}

\newcommand{\bNeVbB}{[\mathrm{Ne}\, {\textsc {vi]}}\, \lambda 24.3\mu \mathrm{m}}

\newcommand{\bOIVb}{[\mathrm{O}\, {\textsc {iv]}}\, \lambda 25.9\mu \mathrm{m}}

\newcommand{\bSIIIbB}{[\mathrm{S}\, {\textsc {iii]}}\, \lambda 33.5\mu \mathrm{m}}

\newcommand{\bSiIIb}{[\mathrm{Si}\, {\textsc {ii]}}\, \lambda 34.8\mu \mathrm{m}}

\newcommand{\bNeIIIbB}{[\mathrm{Ne}\, {\textsc {iii]}}\, \lambda 36.0\mu \mathrm{m}}

\newcommand{\EBV}{E_{\textsc {b-v}}}

\newcommand{\Fnlr}{F_{\textsc {nlr}}}

\newcommand{\lgs}{\log ^{2}S}

\newcommand{\qion}{q_{\mathrm{ion}}}

\newcommand{\tion}{\Delta \theta _{\mathrm{ion}}}

\newcommand{\ml}{m_{\ell }}

\newcommand{\fl}{f_{\ell }}

\newcommand{\kl}{k_{\ell }}

\newcommand{\Eion}{E_{\mathrm{ion}}}

\newcommand{\maxs}{\max S_{\ell }}

\newcommand{\Mo}{M_{\odot }}

\newcommand{\Ro}{R_{\odot }}

\newcommand{\Lo}{L_{\odot }}

\newcommand{\NGC}{\mathrm{NGC}\, 1068}


\title{Infrared spectroscopy of NGC~1068:\\
Probing the obscured ionizing AGN continuum}

\author{Tal Alexander\altaffilmark{1}\\
Dieter Lutz\altaffilmark{2}, Eckhard Sturm\altaffilmark{2}, Reinhard Genzel\altaffilmark{2}\\
Amiel Sternberg\altaffilmark{3}, Hagai Netzer\altaffilmark{3}}

\altaffiltext{1}{Institute for Advanced Study, Olden Lane, Princeton, NJ 08540,USA}

\altaffiltext{2}{Max-Planck-Institut f\"{u}r Extraterrestrische Physik, Postfach 1603, D-85740 Garching, Germany}

\altaffiltext{3}{School of Physics and Astronomy and Wise Observatory, Raymond and Beverly Sackler Faculty of Exact Sciences, Tel Aviv University, Ramat Aviv, Tel Aviv 69978, Israel}

\begin{abstract}
The \emph{ISO-SWS}\footnote{%
Based on observations made with ISO, an ESA project with instruments funded
by ESA member states (especially the PI countries: France, Germany, The Netherlands
and the United Kingdom) and with the participation of ISAS and NASA. The SWS
is a joint project of SRON and MPE. 
} \( 2.5 \)--\( 45\, \mu \mathrm{m} \) infrared spectroscopic observations
of the nucleus of the Seyfert 2 galaxy \( \NGC  \) (see companion paper) are
combined with a compilation of UV to IR narrow emission line data to determine
the spectral energy distribution (SED) of the obscured extreme-UV continuum
that photoionizes the narrow line emitting gas in the active galactic nucleus.
We search a large grid of gas cloud models and SEDs for the combination that
best reproduces the observed line fluxes and NLR geometry. Our best fit model
reproduces the observed line fluxes to better than a factor of 2 on average
and is in general agreement with the observed NLR geometry. It has two gas components
that are consistent with a clumpy distribution of dense outflowing gas in the
center and a more extended distribution of less dense and more clumpy gas farther
out that has no net outflow. The best fit SED has a deep trough at \( \sim \! 4 \)
Ryd, which is consistent with an intrinsic Big Blue Bump that is partially absorbed
by \( \sim \! 6\times 10^{19}\, \mathrm{cm}^{-2} \) of neutral hydrogen interior
to the NLR.
\end{abstract}

\section{Introduction}


The intrinsic spectral energy distribution (SED) of active galactic nuclei (AGN),
which extends from the radio up to \( \gamma  \)-rays, cannot be directly observed
from the Lyman limit and up to several hundred eV due to Galactic and intrinsic
absorption. However, the extreme-UV (EUV) and soft X-ray continuum can be investigated
indirectly by the infrared coronal line emission. These lines are emitted by
collisionally excited forbidden fine-structure transitions of highly ionized
atoms, whose ionization potentials extend well beyond the Lyman limit up to
hundreds of eV. Unlike the strong permitted lines of these ions, which are also
emitted in the obscured EUV, the reddening-insensitive forbidden IR coronal
lines and semi-forbidden optical coronal lines can be observed. Therefore, when
photoionization is the main ionization mechanism, the coronal lines can provide
information on the intrinsic obscured SED and the accretion process that powers
the AGN. This information can be extracted by photoionization models of the
NLR.

The coronal lines are collisionally suppressed in the dense broad line region
(BLR) close to the continuum source and are efficiently emitted only from the
more rarefied gas in the narrow line region (NLR), hundreds of pc away from
the center. It is well established that large quantities of gas attenuate the
continuum emission in many AGN. These gas clouds are detected by narrow UV absorption
lines (e.g. Kriss et al. \cite{Kriss92}) or by X-ray absorption features and
emission lines (e.g. George et al. \cite{George98}). Although their exact location
along the line of sight is unknown, there are reasons to believe that in some
cases they may be inside the NLR. In particular, the warm absorbers that block
the X-ray continuum appear to cover a large fraction of the continuum source
(George et al. \cite{George98}). This raises the possibility that in some AGN
the ionizing SED, which is traced by the coronal lines, is not the intrinsic
one produced by the accretion process, but rather one that is filtered by intervening
absorbers inside the NLR. It has been proposed that such absorbers are common
in Seyfert galaxies, and are responsible for the observed correlations between
the soft X-ray slope and the narrow emission line spectra of Seyfert 1.5 galaxies
(Kraemer, Ruiz \& Crenshaw \cite{KTCG99}).

This study of the Seyfert 2 galaxy \( \NGC  \) is part of the \emph{ISO-SWS}
program on bright galactic nuclei. Previous studies in this program include
the reconstruction of the SED of the Seyfert 2 Circinus galaxy (Moorwood et
al. \cite{Moorwood96}; Alexander et al. \cite{Alexander99}) and of the Seyfert
1 Galaxy NGC~4151 (Alexander et al. \cite{Alexander99}). In both cases we found
evidence of a ``Big Blue Bump'' signature of a thin accretion disk (Shakura
\& Sunyaev \cite{SS73}). However, in the case of NGC~4151 this structure is
masked by a deep absorption trough of an absorber situated between the BLR and
the NLR, which filters the light that photoionizes the NLR. 

In this paper we apply our SED reconstruction method to \( \NGC  \), one of
the closest, brightest and most extensively studied Seyfert 2 galaxies, which
is considered a prototype of this AGN class. The first detection of broad permitted
emission lines in the polarized light of \( \NGC  \) (Antonucci \& Miller \cite{AM85})
provided a major argument for the Seyfert 1 and 2 unification scheme (Antonucci
\cite{Antonucci93}). This scheme postulates that the two Seyfert types have
both broad and narrow line regions and an obscuring torus that lies between
the two. When the torus is face on and the BLR is directly observed, the AGN
is classified as a Seyfert 1. When the BLR is obscured by the torus, the AGN
is classified as a Seyfert 2, and the BLR can be observed only indirectly in
scattered polarized light. The factors that determine the accretion properties
and the ionizing SED of AGN are currently unknown. However, the Seyfert unification
picture implies that all possibly relevant factors being equal, such as luminosity,
host galaxy type or redshift, the intrinsic SED of both AGN types should be
similar. It is therefore of interest to complement our previous study of the
nearby luminous Seyfert 1 galaxy NGC~4151 with a corresponding study of a nearby
luminous Seyfert 2 galaxy with a similar host galaxy type, such as \( \NGC  \).
The \emph{ISO-SWS} observations of \emph{\( \NGC  \)} are presented in a companion
paper (Lutz et al. \cite{Lutz00}) and are used there to derive the gas density
and to place constraints on the structure and dynamics of the NLR.

This paper is organized as follows. In \S\ref{sec:physprop} we summarize the
physical properties of the nucleus of \( \NGC  \) that are needed for constructing
the photoionization models and constraining their results. In \S\ref{s:lines} we
present the emission line flux compilation that we use in our modeling. In \S\ref{s:models}
we briefly discuss the construction and fitting of the NLR photoionization models.
We present the results in \S\ref{s:results} and discuss them in \S\ref{s:discuss}.

\section{The physical properties of NGC~1068}

\label{sec:physprop}

\label{s:n1068}\( \NGC  \) is a barred spiral galaxy at \( z=0.0036 \) (distance
\( D=16.6\,  \)Mpc for \( H_{0}=65\, \mathrm{km}\, \mathrm{s}^{-1}\, \mathrm{Mpc}^{-1} \))
with magnitude \( m_{B}=9.17 \) (e.g. Lipovetsky, Neizvestny \& Neizvestnaya
\cite{LNN88}). Observations of the nucleus of \( \NGC  \) and models of the
nuclear line emission indicate that the gas in the nucleus forms a complex system,
which is composed of various spatial and dynamical components. These are excited
by several physical mechanisms, including photoionization by the nuclear continuum,
photoionization by hot stars, and possibly also by shocks and energetic particles
from a radio jet. In order to isolate the effects of the nuclear continuum and
to construct photoionization models of the NLR it is necessary to understand
the morphology and content of the galactic nucleus. We present here a brief
overview of the properties of the nucleus that are relevant to this work.

\subsection{The galactic nucleus}

\label{s:n1068-nucleus}

The most prominent morphological feature in the nucleus of \( \NGC  \) is the
asymmetric bi-polar pattern of both the radio and the optical line emission.
The radio emission extends over \( \sim \! 15\arcsec  \) and has a sharply
defined northern lobe and a weaker diffuse southern lobe (Wilson \& Ulvestad
\cite{WU83}; Muxlow et al. \cite{Muxlow96}). The southern lobe is both smaller
and redder, which is consistent with the picture that the large northern lobe
is observed above the galactic disk, generally facing the observer, and the
southern lobe is seen through the galactic disk (Unger et al. \cite{ULPA92};
Macchetto et al. \cite{Macchetto94}). Images of the NLR in low excitation lines
show mainly the northern cone (Cecil, Bland \& Tully \cite{CBT90}; Unger et
al. \cite{ULPA92}; Evans et al. \cite{Evans91}). The precise value of the
opening angle associated with the emission maps depends on the way the edge
of the cone is defined and on the assumed position of the nucleus. The location
of the nucleus can be determined to within \( \sim \! 0.05\arcsec  \) by the
center of symmetry of the UV polarization pattern (Capetti et al. \cite{Capetti95},
Kishimoto \cite{Kishimoto99}). The position of the nucleus does not appear
to coincide with the maximum of the continuum emission, which implies that the
nucleus is heavily obscured even in the infrared. This is consistent with an
obscuring torus of column density in excess of \( 10^{24} \) cm\( ^{-2} \),
as is inferred from X-ray (Marshall et al. \cite{Marshal93}) and CO observations
(Tacconi et al. \cite{Tacconi94}). The positioning of the nucleus makes it
possible to estimate the opening angle of the radiation cone at \( \gtrsim 70\arcdeg  \)
up to \( \sim \! 100\arcdeg  \). The ionization cone appears to be only partially
filled.

Marconi et al. (\cite{MWMO96}) find that the coronal line emission peaks \( 0.5\arcsec  \)
NE of the nucleus and extends up to \( \sim \! 4\arcsec  \). They note that
the blueshift of the emission line profile centroid increases, and the FWHM
of the profile decreases with \( \Eion  \) (the ionization energy required
to produce the emitting ion from the preceding ionization stage). They interpret
this as evidence that the emission lines are emitted from outflowing material.
The high ionization lines are emitted in the inner light cone, where the velocity
field is relatively coherent, while the lower ionization lines are emitted from
slower, more extended areas with different velocities. The NLR \( \bOIIIbB  \)
emission extends over the few inner arcseconds (Evans et al. \cite{Evans91};
Unger et al. \cite{ULPA92}; Dietrich \& Wagner \cite{DW98}). The extended
emission line region (EELR) \( \bOIIIbB  \) emission extends over more than
\( 10\arcsec  \) (Unger et al. \cite{ULPA92}). 

The unresolved BLR is seen only in scattered polarized light and has FWHM of
\( \sim \! 3000 \) km~s\( ^{-1} \) (Miller et al \cite{MGM91}). As is seen
in other Seyfert 2 galaxies (Wilson \cite{Wilson88}; Capetti et al. \cite{Capetti96}),
the morphology of the NLR emission maps is correlated with that of the radio
structure (Wilson \& Ulvestad \cite{WU83}; Capetti et al. \cite{CAM97}). In
particular, a 3-dimensional reconstruction of the positions of individual clouds
based on polarimetric measurements also indicates that the northern NLR cone
is directed towards the observer, and the southern part away from the observer
(Kishimoto \cite{Kishimoto99}). The close correspondence between the NLR and
the jet suggests that the jet outflow sweeps and compresses the ambient gas
and thereby increases the line emissivity. The NLR has a very complex structure
(Macchetto et al. \cite{Macchetto94}) and displays large scale clumpiness.
It is composed of many line emitting cloud complexes (Alloin et al. \cite{Alloin83};
Meaburn \& Pedlar \cite{MP1986}; Evans et al. \cite{Evans91}; Dietrich \&
Wagner \cite{DW98}). The individual components have FWHM ranging from \( \sim \! 200 \)
km~s\( ^{-1} \) to \( \sim \! 700 \) km~s\( ^{-1} \), and extend over \( \sim \! 2500 \)
km~s\( ^{-1} \) in velocity space, resulting in an integrated \( \bOIIIbB  \)
profile with FWHM of \( 1150 \) km~s\( ^{-1} \) (Dietrich \& Wagner \cite{DW98}).
The velocity field of the NLR clouds with the lowest FWHM is consistent with
rotation around the nucleus. The bulk velocities of intermediate FWHM clouds
are clearly split relative to the symmetry axis of the radio jet, which suggests
that they are associated with the interaction between the radio jet and the
NLR gas. The highest FWHM clouds are associated with highly polarized emitting
structures (Capetti et al \cite{Capetti95}), and could therefore be a reflected
image of an inner obscured region. The EELR appears to follow the galactic rotation
(Unger et al. \cite{ULPA92}).

The large \emph{ISO} apertures (\( 14\arcsec \times 20\arcsec  \) to \( 20\arcsec \times 33\arcsec  \))
includes both the outflowing inner NLR and the rotating EELR.

\subsection{The line emitting gas}

\label{s:n1068-gas}

There is evidence that at least three different ionization mechanisms are at
work in the NLR. The very high ionization states are probably due to the the
central continuum source. Marconi et al. (\cite{MWMO96}) find that the infrared
coronal line ratios point to photoionization as the main excitation mechanism
of the coronal gas, and that both collisional excitation or photoionization
by very hot stars can be ruled out. At lower ionization states, the jet / ISM
interaction can provide internal sources of excitation in addition to the external
central continuum, for example by fast shocks or cosmic rays. The morphological
connection between the radio and line emission suggests that the jet outflow
shapes the NLR. Estimates of high gas temperatures (Kriss et al. \cite{KDBFL92})
and the existence of dense but highly ionized clouds near the nucleus on both
sides of the jet axis suggest that the jet may also play a role in photoionizing
the clouds (Capetti et al \cite{CAM97}; Axon et al. \cite{Axon98}; Dietrich
\& Wagner \cite{DW98}). Hot stars are a third ionization mechanism. Unresolved
UV continuum point sources in the inner \( 7\arcsec \times 7\arcsec  \), which
are not observed in \( \bOIIIbB  \), could be OB associations (Macchetto et
al. \cite{Macchetto94}). Hot stars are certainly a component in the ring-like
structure that surrounds the nucleus. An ellipse of H\,\textsc{ii} emission
delineates the NLR at an average radius of \( 13\arcsec  \) (Cecil, Bland \&
Tully \cite{CBT90}; Bruhweiler et al. \cite{BTA91}) and starburst knots and
CO emission encircle the nucleus and the NLR at an average radius of \( 18\arcsec  \)
(Planesas, Scoville \& Myers \cite{Planesas91}). However, the overall similarity
in the profiles of the mid-infrared high-excitation lines, which cannot be excited
by hot stars, and the profiles of the intermediate excitation lines (\( \Eion >30\, \mathrm{eV} \)),
which could be excited by hot stars, strongly suggests that gas excited by hot
stars within the large \emph{ISO} aperture does not contribute more than \( \sim \! 20\% \)
to these NLR line fluxes (Lutz et al. \cite{Lutz00}). An additional complication
due to stars is contamination of the observed emission line spectrum with stellar
absorption features. This appears in the difference spectrum of the \( 30\arcsec \oslash  \)
and \( 18\arcsec \oslash  \) \emph{HUT} apertures, which shows both a reddened
early-type stellar continuum and stellar absorption features (Kriss et al. \cite{KDBFL92}).

As is discussed by Lutz et al. (\cite{Lutz00}), the \emph{ISO-SWS} line ratios
indicate that the mid-IR lines are emitted from gas with a hydrogen density
of \( \sim \! 2000\, \mathrm{cm}^{-3} \). The NLR appears to contain also higher
density gas. The density of individual knots in the high ionization core is
estimated at \( 10^{4} \) to \( 3\times 10^{4} \) cm\( ^{-3} \) from \emph{HST}
measurements of the \( [\textrm{S}\, \textsc {ii}] \) doublet, while the overall
density \( \sim \! 1\arcsec  \) from the nucleus varies between \( 10^{3} \)
to \( 4\times 10^{3} \) cm\( ^{-3} \) (Capetti et al. \cite{CAM97}).

The ionization parameter\footnote{%
\( U\equiv Q_{\mathrm{ion}}/4\pi r^{2}nc, \) where \( Q_{\mathrm{ion}} \)
is the ionizing photon luminosity, \( n \) is the hydrogen density at the illuminated
face of the cloud, \( r \) is the distance of the face of the cloud from the
continuum source and \( c \) is the speed of light. The local ionization parameter
in the cloud falls with increasing depth due to absorption and geometrical dilution
of the radiation field.
} also appears to vary across the NLR. The \( \bOIIIbB /(\Ha +[\mathrm{N}\, {\textsc {ii]}}\, \lambda 6584) \)
ratio (Capetti et al. \cite{CAM97}) traces a high excitation core (\( \log U\sim -2.5 \))
in the inner \( 1\arcsec \times 2\arcsec  \) north of the nucleus, followed
by a lowered ionization halo (\( \log U\sim -3.3 \)) out to \( \sim 4\arcsec  \)
from the nucleus, and then intermediate ionization filaments (\( \log U\sim -2.8 \))
which extend out to the EELR. 

Estimates of dust reddening in the NLR of \( \NGC  \) range from \( \EBV =0.07 \)
for the continuum (Kriss et al \cite{KDBFL92}), to 0.20 (Marconi et al \cite{MWMO96}),
0.40 (Shields \& Oke \cite{SO75}; Neugebauer et al \cite{Neugebauer80}; Ward
\cite{Ward87}), and \( \EBV =0.52 \) (Koski \cite{Koski78}) for the NLR.
The analysis of Kraemer et al. (\cite{KRC98}) indicates that there may be some
dust mixed with gas in varying amounts. The observed line ratios in the NLR
suggest that the O/N abundance is less than solar. Netzer (\cite{Netzer97})
and Netzer \& Turner (\cite{NT97}) interpret this as an indication of under-abundant
oxygen (see also Sternberg, Genzel \& Tacconi \cite{SGT94}), and propose that
the abundances of He:C:N:O:Ne:Mg:Si:S:Ar:Fe relative to hydrogen are \( (100:3.7:1.1:2.7:1.1:0.37:0.35:0.16:0.037:0.4)\times 10^{-4} \),
respectively. Kraemer et al. (\cite{KRC98}) interpret the line ratios as showing
an overabundance of nitrogen, as well as hinting at higher than solar iron and
neon, and propose abundance ratios of \( (100:3.4:3.6:6.8:2.2:0.33:0.31:0.15:0.037:0.8)\times 10^{-4} \).

Finally, Kraemer et al. (\cite{KRC98}) model the NLR emission with a multi-component
gas model, and suggest that the low ionization emission lines are emitted by
gas that is partially screened by a dense, optically thin component. They use
their models to estimate that the filling factor is \( F\sim 10^{-4} \).

\subsection{The ionizing continuum}

\label{s:n1068-SED}

Unlike the situation in Seyfert 1 galaxies, where it is possible to observe
the intrinsic SED outside the obscured range, the intrinsic SED of \( \NGC  \)
cannot be directly observed even in the optical or X-ray bands. Only a small
fraction of the AGN light is scattered into the line of sight, and can be observed
against the host galaxy in polarized light. Pier et al (\cite{Pier94}) compiled
various continuum measurements in the optical, UV and X-ray, and carefully took
account of aperture differences, star-light contamination, reflection by dust
and bremsstrahlung emission from the scattering plasma. They conclude that the
resulting reflected SED is broadly similar to that observed in Seyfert 1 galaxies.
Pier et al (\cite{Pier94}) also list various estimates of the fraction of light
reflected by the scatterer. These values range from \( f_{\textrm{refl}}=10^{-3} \)
(Bland-Hawthorn \& Voit \cite{BH93}) to \( \sim \! 0.05 \) (Bland-Hawthorn,
Sokoloski \& Cecil \cite{BSC91}). Pier et al (\cite{Pier94}) argue that the
most reliable estimate is \( f_{\textrm{refl}}\sim 0.01 \) to within a factor
of a few. Because the reflectors are more than a hundred light years away from
the continuum source (Miller et al. \cite{MGM91}), short-term continuum variability
is unlikely to affect the reconstruction of the reflected SED.

We adopt the Pier et al. nuclear continuum SED in the UV and X-ray as the template
SED (Fig.~\ref{fig:temp}), and further extend it from 10~keV to 100~keV with
a slope of \( F_{\nu }\propto \nu ^{-1} \). We enumerate on the unobserved
UV to soft X-ray range to find the best fitting SED.

\section{The observed line flux compilation}

\label{s:lines}

In addition to the ISO-SWS mid-IR lines fluxes presented in Lutz et al. (\cite{Lutz00}),
we compiled a list of UV to IR lines from the literature. The compilation initially
included a list of \( \sim \! 120 \) measured line fluxes, which were obtained
over the last \( \sim \! 30 \) years using various instruments with different
apertures, spectral resolutions and reduction techniques. The observed lines
were taken from the following references, listed by spectral band with the aperture
used (where given): UV lines from Kriss et al (\cite{KDBFL92}) (\( 18\arcsec \oslash  \));
optical lines from Osterbrock \& Parker (\cite{OP65}) (trailing long slit),
Anderson (\cite{Anderson70}) (\( 8\arcsec \times 8\arcsec  \)), Wampler (\cite{Wampler71})
(\( 10\arcsec \oslash  \)), Koski (\cite{Koski78}) (\( 2.7\arcsec \times 3.4\arcsec  \))
and Neugebauer et al (\cite{Neugebauer80}) (\( \sim \! 10\arcsec \times 20\arcsec  \));
optical to near-IR lines from Shields \& Oke (\cite{SO75}) (\( 10\arcsec \times 10\arcsec  \)),
near-IR lines from Oliva \& Moorwood (\cite{OM90}) (\( 6\arcsec \times 6\arcsec  \)),
Marconi et al (\cite{MWMO96}) (long slit of width \( 4.4\arcsec  \)), Osterbrock
\& Fulbright (\cite{OF96}) (\( 3\arcsec \times 18\arcsec  \)) and Osterbrock,
Tran \& Veilleux (\cite{OTV92}) (long slit of width \( 1.2\arcsec  \)); IR
lines from Thompson (\cite{Thompson96}) (\( 2\arcsec \times 10\arcsec  \))
and the ISO-SWS (\( 14\arcsec \times 20\arcsec  \) to \( 20\arcsec \times 33\arcsec  \)).

As is discussed in detail by Alexander et al. (\cite{Alexander99}), the result
of such a compilation is generally not self-consistent, and only a small subset
of the of the \( \sim \! 120 \) lines can be used. First, we excluded low and
medium excitation lines (\( \Eion \lesssim 100\,  \)eV) observed through apertures
whose smaller dimension is less than \( 3\arcsec  \), so as to avoid significant
loss of light due to incomplete coverage of the line emitting region. The observed
emission from the high excitation lines is centrally concentrated in the inner
\( <4\arcsec  \) (Marconi et al \cite{MWMO96}), which are covered even by
the smallest apertures used. Second, we excluded lines with \( \Eion <30\,  \)eV
to avoid using lines that may be photoionized primarily by young hot stars or
other non-AGN, lower-energy excitation processes. This also reduces the loss
of light bias. Third, we excluded narrow lines whose measured flux is uncertain,
either because the flux is very low (flux less than \( 5\times 10^{-13}\,  \)erg
s\( ^{-1} \) cm\( ^{-2} \), \( \sim \! 2\% \) of the strongest line), or
because it has a significant broad component, such as the \( \NV  \) and \( \CIV  \)
lines. Fourth, we excluded the \( \bFeXb  \) and \( \bFeXIb  \) lines, whose
collision strengths are highly uncertain and therefore cannot be modeled reliably.
The final, much reduced line list includes 22 lines (Table~\ref{tbl:flux}),
which we use for obtaining the best-fit SED. Whenever more than one measurement
of the line exists, we quote the average flux and use the rms scatter as an
error estimate. Important IR lines that were not included in the final list
were nevertheless compared to the best fit model predictions to verify that
there are no gross inconsistencies.

\section{The photoionization models}

\label{s:models} 

The method of constructing photoionization models for the NLR and, in particular,
the ``\( \lgs  \) fit procedure'' for obtaining the best fitting SED and
gas parameters is described in detail in Alexander et al. (\cite{Alexander99}),
and is summarized here briefly. We parameterize the SED as a piece-wise broken
power-law (Fig.~\ref{fig:temp}), and enumerate on the different possibilities
of connecting the power-law segments. We test a large number of simplified NLR
gas models, which consist of optically thick (radiation bounded) clouds whose
ionized surfaces partially cover a spherical shell around the continuum source.
Each cloud extends in the radial direction as far as it takes to effectively
absorb all the ionizing UV photons (specifically, until the hydrogen ionization
fraction falls below 2\%). The gas clouds are parameterized by their chemical
composition, the hydrogen density \( n \), the ionization parameter \( U \)
at the irradiated face of the cloud, and the filling factor \( F \). In addition,
an asymmetry parameter \( A \) expresses the ratio between the ionizing flux
directed towards the NLR and that directed towards the observer. This describes
situations where the continuum source is not isotropic, or where only a fraction
\( f_{\mathrm{refl}} \) of the continuum is reflected towards the observer
(\( A=1/f_{\mathrm{refl}} \)).

For each NLR gas model, the fit procedure uses the observed line fluxes to derive
the best-fit SED for that gas model, and from it to derive in a self-consistent
way the corresponding covering factor \( C, \) the inner NLR angular radius
\( \theta  \), the width of the ionized region \( \tion  \) (defined here
as the radial extent of the Balmer lines emitting gas) and the reddening coefficient
\( \EBV  \). These parameters are constrained by the observations and so can
be used to limit the range of acceptable NLR gas models. The \( \lgs  \) fit
procedure assigns a score \( S \) to the best-fit model, which means that the
model line fluxes fit the observed ones up to a factor \( S \), on average.
The worst-fitting line and the factor by which it deviates from the observed
value are also recorded. Monte-Carlo simulations are used to calculate confidence
limits on the best-fit SED. In addition, we calculate the correlation between
the model-to-data line ratios and the lines wavelengths, ionization potentials,
critical densities and ``deplitivity'' (the tendency of an element to be depleted
into dust). These residual correlations test whether the remaining inconsistencies
in the best-fit model are related to inaccurate modeling of the reddening, the
spectral hardness / softness of the continuum, the gas density or its dust content.
The correct model should not display any such correlations. The final, global
best-fit SED is the one with the best \( S \)-score among all the NLR gas models
that are consistent with the observed NLR geometry and reddening, whose worst-fitting
line is not too far from the observed value, and which display no significant
residual correlations.

We assume in all models that the gas clouds have constant density and that \( f_{\mathrm{refl}}=0.005 \).
We investigate two classes of models. The first consists of models with a single
type of cloud. We enumerate on different values for the ionization parameter
(\( \log U=-1,-2 \) and \( -3 \)), gas density (\( n=2000 \) and \( 10^{4}\, \mathrm{cm}^{-3} \))
and filling factor (\( \log F=-2,-3 \) and \( -4 \) at the ionized surface).
We test two different sets of non-solar abundances, the low oxygen set and high
nitrogen set (\S\ref{s:n1068-gas}). We test two different possibilities for
the radial run of the filling factor. The first is that \( F \) is constant,
which corresponds to either a static distribution of clouds, a rotating distribution
of clouds, a linear constant velocity outflow, a strongly decelerating outflow
at a constant opening angle, or an outflow where clouds are continuously added
to the flow (e.g. from the molecular torus). The second is \( F\propto r^{-2} \),
which corresponds to a constant velocity and constant opening angle outflow
where the clouds are formed at the base of the flow. The second class of models
have two types of gas clouds, and are constructed by combining all the possible
pairs of one component models. We assume that the clouds do not obscure one
another. 

The photoionization calculations were carried out using the numerical photoionization
code \noun{ion9}9, the 1999 version of the code \noun{ion} described by
Netzer (\cite{Netzer96}).

\section{Results}

\label{s:results}

\subsection{Single component models}

We find that single component models generally fail to fit the observed line
ratios and the observed constraints on the geometry of the NLR. Low \( U \)
models (\( \log U=-3 \)) cannot reproduce the high excitation lines regardless
of the values of the other model parameters. For the \( F\propto r^{-2} \)
models, the low \( U \) models significantly over-estimate the size of the
NLR, while the high-\( U \) models require unphysical covering factors exceeding
unity. Single component models with a high \( U \) and low constant filling
factor do somewhat better, although such models are problematic because a constant
filling factor is inconsistent with simple scenarios of NLR outflow. The best
fit single component model has low oxygen abundance with \( \log U=-1 \), \( \log n=3.3 \)
and \( \log F=-3 \). This model can reproduce the observed lines up to a factor
of 2 and predicts \( \theta =0.8\arcsec  \), \( \tion =4.8\arcsec  \), \( C=0.35 \)
and \( \EBV =0.18. \) These values are roughly consistent with the observed
constraints. However, this model under-predicts the observed \( \OVI  \) flux
by a factor of \( \sim \! 5 \) and shows a residual negative correlation with
\( \lambda _{0} \) at the 5\% confidence level. Like all the models investigated
here, the best fit SED displays a deep trough at 4 Ryd (\( \log f=-27.4,\, -29.0,\, -27.4,\, -28.2 \)
at 2, 4, 8 and 16 Ryd, respectively). A fit of similar quality is obtained with
the high nitrogen abundance set.

\subsection{Two component models}

A better fit to the observations is provided by the best-fit two component model
(Table~\ref{tbl:fit}). This model fits the 22 observed line fluxes to within
a factor of 1.9 on average (experience shows this is as well as one can expect
for AGN photoionization models). The worst fitting line, \( \bArVIb  \), is
under-predicted by a factor of 4. The model-to-data ratios of individual lines
are displayed in Fig.~\ref{fig:ratios}. The low excitation IR lines (\( \Eion <30\, \mathrm{eV} \))
such as \( \bNeIIb  \), \( \bSIIIbA  \) and \( \bSIIIbB  \), which were not
used in the fit, are nevertheless all consistent with the observations to within
a factor of 2. However, the model-to-data line ratios for these lines are not
much smaller than 1, as would be expected if there is a significant contamination
from gas photoionized by starbursts. The high excitation IR lines such as \( [\mathrm{S}\, {\textsc {ix]}}\, \lambda 1.25\mu \mathrm{m} \),
\( [\mathrm{Si}\, {\textsc {x]}}\, \lambda 1.4\mu \mathrm{m} \) and \( [\mathrm{Si}\, {\textsc {ix]}}\, \lambda 2.6\mu \mathrm{m} \),
which were not used in the fit, are also consistent with the observations to
within a factor of 3. There are no statistically significant residual correlations
between these ratios and \( \lambda _{0} \), \( \Eion  \), the deplitivity
or \( n_{c} \). The best fit values of the extinction \( (\EBV \sim 0.2 \))
is in agreement with other estimates (\S\ref{s:n1068-gas}). The covering factors
are somewhat larger than is indicated by the observed opening angle of the emission
cone (\S\ref{s:n1068-nucleus}). The dense compact component requires a covering
factor of \( C=0.45 \), which corresponds to a bi-cone with an opening angle
of \( \sim \! 115\arcdeg  \). The less dense, more extended component requires
a covering factor \( C=0.26 \), which corresponds to a single cone with an
opening angle of \( \sim \! 130\arcdeg  \) or a bi-cone with opening angle
of \( \sim \! 90\arcdeg  \). Figure~\ref{fig:mix} shows the best fit SED for
this gas model together with the 99.9\% confidence limits on it (the lower confidence
limits at 2, 4 and 16 Ryd were not calculated as they extend beyond the SED
grid). This SED shows the generic trough that is seen in the best-fit SED of
all the models we investigated. This model has a less than solar oxygen abundance
(Netzer \cite{Netzer97}; Netzer \& Turner \cite{NT97}). We find only a few
two-component models with high nitrogen abundance that come close to a reasonable
fit to the observations. Of these, almost all display a negative correlation
with deplitivity at the 5\% to 10\% significance, which may indicate that the
line emitting gas is more depleted than is assumed by the high nitrogen abundance
set.

We conclude that best-fit two component model provides a better, but not an
overwhelmingly better fit to the observed line ratios than the single component
model. Its marked advantage over the single component models lies in its consistency
with the observed NLR geometry and kinematics. This is further discussed in
the next section.

\section{Discussion}

\label{s:discuss}

As was discussed in detail by Alexander et al. (\cite{Alexander99}), there
are various degeneracies between the parameters that describe the gas model
(\( n, \) \( A \), \( F \), \( U \), \( C \), \( \theta  \), and \( \tion  \)).
These degeneracies allow the fit procedure to converge to a robust best-fit
SED even when the assumed values of \( n \), \( A \), \( F \) or \( U \)
significantly differ from their true values, since this can be compensated to
a large degree by a suitable modification of the gas geometry (\( C \), \( \theta  \),
and \( \tion  \)). This property of the fit procedure is especially important
in the analysis of \( \NGC  \), where observations indicate that the actual
properties of the NLR gas are much more complex than can be modeled by our family
of simplified gas models. For this reason, we place more weight on the fact
that all the best-fit SED models, whether one or two-component, display a deep
trough at 4 Ryd than on the determination of the exact values of the gas parameters. 

The trough in the SED is required for reproducing the relative line fluxes of
the high and low ionization species. To check the robustness of this result,
we attempted to re-fit the observed line fluxes with all of our one and two-component
models using an approximate single power-law SED (\( \log f=-25.8,\, -26.6,\, -27.4,\, -28.2 \)
at 2, 4, 8 and 16 Ryd, respectively), which was held fixed in the fit procedure.
In all cases the low excitation lines were over-estimated with respect to the
high excitation lines, regardless of the values of \( U \), \( n \) or \( F \).
For example, when the power-law SED is applied to the best-fit two-component
gas model (Table~\ref{tbl:fit}), the low excitation lines (\( \Eion \lesssim 50\, \mathrm{eV} \))
are over-estimated by the model by up to a factor of 13, whereas the high excitation
lines (\( \Eion \gtrsim 100\, \mathrm{eV} \)) are under-estimated by up to
a factor of 23. The overall mismatch of this SED with the observed line fluxes
is reflected in both the poor fit score of \( S=4.1 \) and in the very strong
residual anti-correlation between the line ratios and \( \Eion  \), whose random
probability is \( 10^{-4} \). We caution against generalizing this result to
mean that every AGN that exhibits a hard emission line spectrum has an absorbed
ionizing SED. The emission line spectrum reflects the gas parameters, such as
\( U \), \( n \) and \( F \), no less than it does the ionizing SED. It is
necessary to have some knowledge of the likely range of values for these parameters
in order to interpret the hardness of the line spectrum. Alexander et al. (\cite{Alexander99})
provide a counter-example where subtle cancellations between the SED and the
gas properties lead to a situation where an AGN (NGC4151) with a hard absorbed
SED has a \emph{softer} emission line spectrum than another AGN (Circinus) with
a soft unabsorbed Big Blue Bump. 

Although we do not claim to fix the gas parameters with certainty, the best
fit model (Table~\ref{tbl:fit}) is reassuringly consistent with the observations,
which broadly indicate that the integrated NLR emission originates in two components.
Component A of the model can be interpreted as a system of dense (\( n=10^{4} \)~\( \mathrm{cm}^{-3} \)
), centrally concentrated (\( 0.3\arcsec <\theta <1.4\arcsec  \)) outflowing
gas clouds (\( F\propto r^{-2} \)) with a relatively high filling factor (\( F\sim 0.01 \))
and high ionization parameter \( (\log U=-1 \)). Component B of the model can
be interpreted as a more extended distribution (\( 1.9\arcsec <\theta <4.5\arcsec  \))
of lower density gas (\( n=2\times 10^{3} \)~\( \mathrm{cm}^{-3} \)) with
no net outflow (\( F=\mathrm{const}. \)), with a lower filling factor (\( F=0.001) \)
and a lower ionization parameter (\( \log U=-2 \)). Component A contributes
58\% of the total line flux in 22 lines listed in Table~\ref{tbl:flux}, with
the contribution to individual lines ranging from 45\% of the low excitation
\( \bSIVb  \) (\( \Eion =34.8\, \mathrm{eV} \)) to more than 99.9\% of the
very high excitation line \( \bSiIXbB  \) (\( \Eion =303.2\, \mathrm{eV} \)).
Its large covering factor indicates that there is probably a significant contribution
of flux from the inner \( \sim \! 1\arcsec  \) of the diffuse SW emission cone
as well as from the bright NE cone. Component B contributes the remaining 42\%
of the total line flux, mainly in the lower excitation lines. Its covering factor
is small enough for it to be concentrated mostly in the NE bright emission cone,
as is observed. 

The best-fit procedure indicates that models with the low oxygen abundance set
fit the observed line fluxes somewhat better than models with the high nitrogen
abundance set. In particular, the \( \mathrm{O}\, {\textsc {iii]}}\, \lambda 1663 \),
whose unusual relative weakness was an important argument for assuming non-solar
abundances (Netzer \cite{Netzer97}; Kraemer et al. \cite{KRC98}), is well
reproduced by the best-fit low oxygen model with a model-to-data line flux ratio
of 1.4 (Fig.~\ref{fig:ratios}), even though it was not included in our fit
since it didn't pass the minimal flux criterion. 

The trough in the best fit SED (Fig.~\ref{fig:mix}) can be interpreted as an
absorption trough due to an absorber between the continuum source and the NLR.
The energy resolution of the SED template, which is limited by the computational
cost of enumerating over all the SED combinations, is too low to allow detailed
modeling of the absorber. Figure~\ref{fig:abs_sed} shows an example of how
a quasi-thermal big blue bump that is absorbed by neutral hydrogen would appear
in our low resolution SED reconstruction. We find that the trough is consistent,
for example, with an absorber that shadows the entire NLR (\( C_{\mathrm{abs}}=1) \)
and has a column density of \( N_{H^{0}}=6\times 10^{19}\, \mathrm{cm}^{-2} \)
in neutral hydrogen, or with an absorber that allows a small leakage of unfiltered
radiation (\( C_{\mathrm{abs}}=0.999 \)) and a column density of \( N_{H^{0}}=10^{20}\, \mathrm{cm}^{-2} \).
A similar trough, consistent with an absorber of \( C_{\mathrm{abs}}>0.99 \)
and \( N_{H^{0}}=5\times 10^{19}\, \mathrm{cm}^{-2} \), was discovered in the
reconstructed SED of Seyfert 1 galaxy NGC~4151 (Alexander et al. \cite{Alexander99}).
That absorber was also detected in the \emph{HUT} absorption line spectra of
the UV continuum of NGC~4151 (Kriss et al. \cite{Kriss92}, \cite{Kriss95})
. The \emph{HUT} spectra of \( \NGC  \) (Kriss et al \cite{KDBFL92}) do not
have a high enough S/N to allow the detection of absorption lines in the scattered
UV continuum of this AGN or against the stellar background (G. Kriss, private
comm.). We predict that future sensitive absorption line studies should reveal
the presence of such an absorber. 

The bias in our results due to the fact that we neglected the line emission
from the absorbing gas is likely to be small if the absorber is similar to the
dense, high velocity UV absorber that was detected in NGC~4151. Such an absorber
will not emit forbidden lines, and its permitted lines will be broader than
typical NLR lines. Only 3 of the 22 lines we used in our fit are permitted lines,
and we did not use lines that are contaminated by broad components. A highly
ionized and optically thin absorber will produce strong \( \OVI  \) line emission
in excess of the typical NLR emission. It is therefore interesting that unlike
the two forbidden \( \bOIIIbB  \) and \( \bOIVb  \) lines and the semi-forbidden
\( \OIIIb  \) line, which are well reproduced by the best fit model, the observed
\( \OVI  \) line is 3.2 stronger than predicted (Fig.~\ref{fig:ratios}). 

We have, up to now, applied our SED reconstruction method to \emph{ISO-SWS}
observations of IR coronal lines of three AGN: the Seyfert 2 Circinus galaxy
(Moorwood et al. \cite{Moorwood96}; Alexander et al. \cite{Alexander99}),
the Seyfert 1 galaxy NGC~4151 (Alexander et al. \cite{Alexander99}), and the
Seyfert 2 galaxy \( \NGC  \) (this work). In one of these (Circinus), we detect
a Big Blue Bump that peaks at \( \gtrsim 50\, \mathrm{eV} \). In the other
two we detect deep troughs, which are consistent (but not exclusively so) with
a Big Blue Bump that is absorbed by neutral gas interior to the NLR. Our findings
thus far are consistent with the picture that luminous Seyfert galaxies are
powered by thin accretion disks that produce a quasi-thermal Big Blue Bump,
and that in a large fraction of them the NLR sees a partially absorbed ionizing
continuum, as suggested by Kraemer et al. (\cite{KRC98}).

\acknowledgments

This work was supported by DARA under grants 50-QI-8610-8 and 50-QI-9492-3,
and by the German-Israeli Foundation under grant I-0551-186.07/97.

\clearpage

\figcaption[sed_template.eps]{The template used for enumerating on the SED of
\( \NGC  \). The most and least luminous SEDs are indicated by the lines, together with
the relative sense of their spectral hardness between 1 and 150 Ryd. \label{fig:temp}}

\figcaption[best_mix.eps]{The best fit SED of the two-component model
(Table~\ref{tbl:fit}). \label{fig:mix}}

\figcaption[lratios.eps]{The model to data line ratios for the best fit
two-component model (Table~\ref{tbl:fit}). The $\OIIIb$ line ratio (white
circle), which was not used in the fitting procedure is also
displayed. \label{fig:ratios}}

\figcaption[abs_sed.eps]{An example of how 
a quasi-thermal Big Blue Bump that is absorbed by neutral hydrogen
would appear in the low resolution reconstructed SED. The unabsorbed
bump (thin line) and the bump after absorption by a $C_{\rm abs}=1$,
$N_{H^0}=6\times10^{19}$\,cm$^{-2}$ absorber (long dashed line) and a
$C_{\rm abs}=0.999$, $N_{H^0}=10^{20}$\,cm$^{-2}$ absorber (short
dashed line) are superimposed on the best fit SED
model.\label{fig:abs_sed}}

\clearpage

\begin{deluxetable}{lr@{.}lr@{.}lr@{.}lr@{.}l} \small
\tablecaption{The compiled emission line flux list. \label{tbl:flux}}
\tablewidth{0pt}
\tablehead{
\colhead{Line} &
\multicolumn{2}{c}{\(\lambda_{0}\)}&
\multicolumn{2}{c}{\(E_{\textrm{ion}}\)\tablenotemark{a}}&
\multicolumn{2}{c}{\(f_{\ell}\)\tablenotemark{b}}&
\multicolumn{2}{c}{\(\Delta f_{\ell}\)\tablenotemark{c}}\\ 
\colhead{}&
\multicolumn{2}{c}{\(\mu\)m}&
\multicolumn{2}{c}{eV}&
\multicolumn{4}{r}{\(10^{-13}\,{\textrm{erg}}\,{\textrm{s}}^{-1}{\textrm{cm}}^{-2}\)}
}
\tablecolumns{9}
\startdata
\cutinhead{UV, optical and NIR lines}
O\,{\textsc{vi}}& 0&1032+1038 & 113&9 & 37&4& 3&1\nl
N\,{\textsc{iv}}]& 0&1487 & 47&4 & 5&1 & 1&1\nl
 He\,{\textsc{ii}}&    0&1640 &  54&4 &  17&7     & 5&2\nl
[Ne\,{\textsc{v}}]&    0&3426 &  97&1 &  15&7     & 6&7\nl
[Ne\,{\textsc{iii}}]&  0&3868+3969 &  41&0 &  19&2     & 2&6\nl
 He\,{\textsc{ii}}&    0&4686 &  54&4 &  6&15     & 1&50\nl
[O\,{\textsc{iii}}]&   0&4959+5007 &  35&1 &  256&     & 27&\nl
[Si\,{\textsc{vi}}]&   1&96   & 166&8 &  7&5      & 0&5\nl
[Si\,{\textsc{vii}}]&  2&48   & 205&1 &  8&3      & 0&9\nl
\cutinhead{ISO-SWS IR lines}
[Mg\,{\textsc{viii}}]&  3&028   & 224&9 &  11&      & 1&1\nl
[Si\,{\textsc{ix}}]&    3&936   & 303&2 &  5&0      & 0&6\nl
[Mg\,{\textsc{iv}}]&    4&49    &  80&1 &  7&6      & 1&5\nl
[Ar\,{\textsc{vi}}]&    4&528   &  75&0 &  15&      & 3&\nl
[Mg\,{\textsc{vii}}]&   5&50    & 186&5 &  13&      & \multicolumn{2}{l}{?}\nl
[Mg\,{\textsc{v}}]&     5&61    & 109&2 &  18&      & 2&\nl
[Ne\,{\textsc{vi}}]&    7&64    & 126&2 &  110&     & 11&\nl
[S\,{\textsc{iv}}]&    10&51    &  34&8 &  58&      & 6&\nl
[Ne\,{\textsc{v}}]&    14&32    &  97&1 &  97&0     & 9&7\nl
[Ne\,{\textsc{iii}}]&  15&56    &  41&0 &  160&     & 32&\nl
[Ne\,{\textsc{v}}]&    24&32    &  97&1 &  70&      & 7&\nl
[O\,{\textsc{iv}}]&    25&89    &  54&9 &  190&     & 20&\nl
[Ne\,{\textsc{iii}}]&  36&01    &  41&0 &  17&      & 3&\nl
\enddata
\tablenotetext{a}{
The ionization energy required to produce the emitting ion from the
preceding ionization stage.}
\tablenotetext{b}{
The observed flux. For permitted lines, the flux of the decomposed narrow
component is quoted.}
\tablenotetext{c}{
The error estimate on the observed flux. The ISO-SWS errors are estimated by
the scatter in various methods for defining the underlying continuum and
measuring the line. The errors do not include systematic calibration
errors, which are generally smaller than 30\%. A question mark means that
error estimates are unavailable.}
\end{deluxetable}

\clearpage
\begin{deluxetable}{llcc}
\small
\tablecaption{The best-fit two-component model. \label{tbl:fit}}
\tablewidth{5in}
\tablehead{
\colhead{} & \colhead{} & \colhead{Component A} & \colhead{Component B}
}
\tablecolumns{4}
\startdata
{Input parameters:}&&&\nl
&$U$          &  0.1                  &  0.01                \nl
&$n$          &$10^4$ cm$^{-3}$      &$2\times10^3$ cm$^{-3}$\nl              
&Composition  & low oxygen            & low oxygen           \nl
&$A$          &  $200$                &  $200$               \nl
&$F$\tablenotemark{a}&
                $10^{-2}h$           &  $10^{-3}h$           \nl
&$\case{{d\log F}}{{d\log r}}$
             &  $-2$                 &  $0$                  \nl
{Best-fit results:\tablenotemark{b}}&&&\nl
&$S$          &  \multicolumn{2}{c}{$1.9$}                   \nl
&$\max S_\ell$\tablenotemark{c}
              &  \multicolumn{2}{c}{4.0$^{-1}$}               \nl
&worst line
              &  \multicolumn{2}{c}{$\bArVIb$}                \nl
&$\log f_2$\tablenotemark{d}
              &  \multicolumn{2}{c}{-29.0 (-29.0, -26.6)}     \nl
&$\log f_4$   &  \multicolumn{2}{c}{-29.0 (-29.0, -26.6)}    \nl
&$\log f_8$   &  \multicolumn{2}{c}{-27.4 (-27.4, -26.6)}    \nl
&$\log f_{16}$&  \multicolumn{2}{c}{-29.0 (-29.0, -28.2)}    \nl
&$C$          &  0.45                 &  0.29                \nl
&$\theta$     &  $0.26\arcsec$        &  $1.9\arcsec$        \nl
&$\tion$      &  $1.1\arcsec$         &  $2.6\arcsec$        \nl
&$\EBV$       &  \multicolumn{2}{c}{0.23 (0.18, 0.29)}       \nl
&$\qion$      &  \multicolumn{2}{c}{0.24 s$^{-1}$\,cm$^{-2}$}\nl
&$\log Q_{\rm ion}$/s$^{-1}$ \tablenotemark{e}&
                \multicolumn{2}{c}{54.2}                     \nl
&$\langle h\nu \rangle$\tablenotemark{f}&
                 \multicolumn{2}{c}{5.1 Ryd}                 \nl
{Residual correlations:}&& Correlation & Random prob.        \nl
&$\lambda_0$  & $-0.19$ & $0.23$\nl
&$\Eion$      & $ 0.00$ & $1.00$\nl
&depletion    & $+0.11$ & $0.70$\nl
&$n_c$        & $+0.03$ & $0.86$\nl
\enddata
\tablenotetext{a}{Assuming $h=0.65$.}
\tablenotetext{b}{Values in parentheses are the 99.9\% confidence intervals.}
\tablenotetext{c}{Model to data ratio, given as the reciprocal when $<1$.}
\tablenotetext{d}{Flux in erg\,s$^{-1}$\,cm$^{-2}$\,Hz$^{-1}$.}
\tablenotetext{e}{Ionizing photon luminosity assuming isotopic emission.}
\tablenotetext{f}{Mean ionizing photon energy between 1 and 16 Ryd.}
\end{deluxetable}

%
%



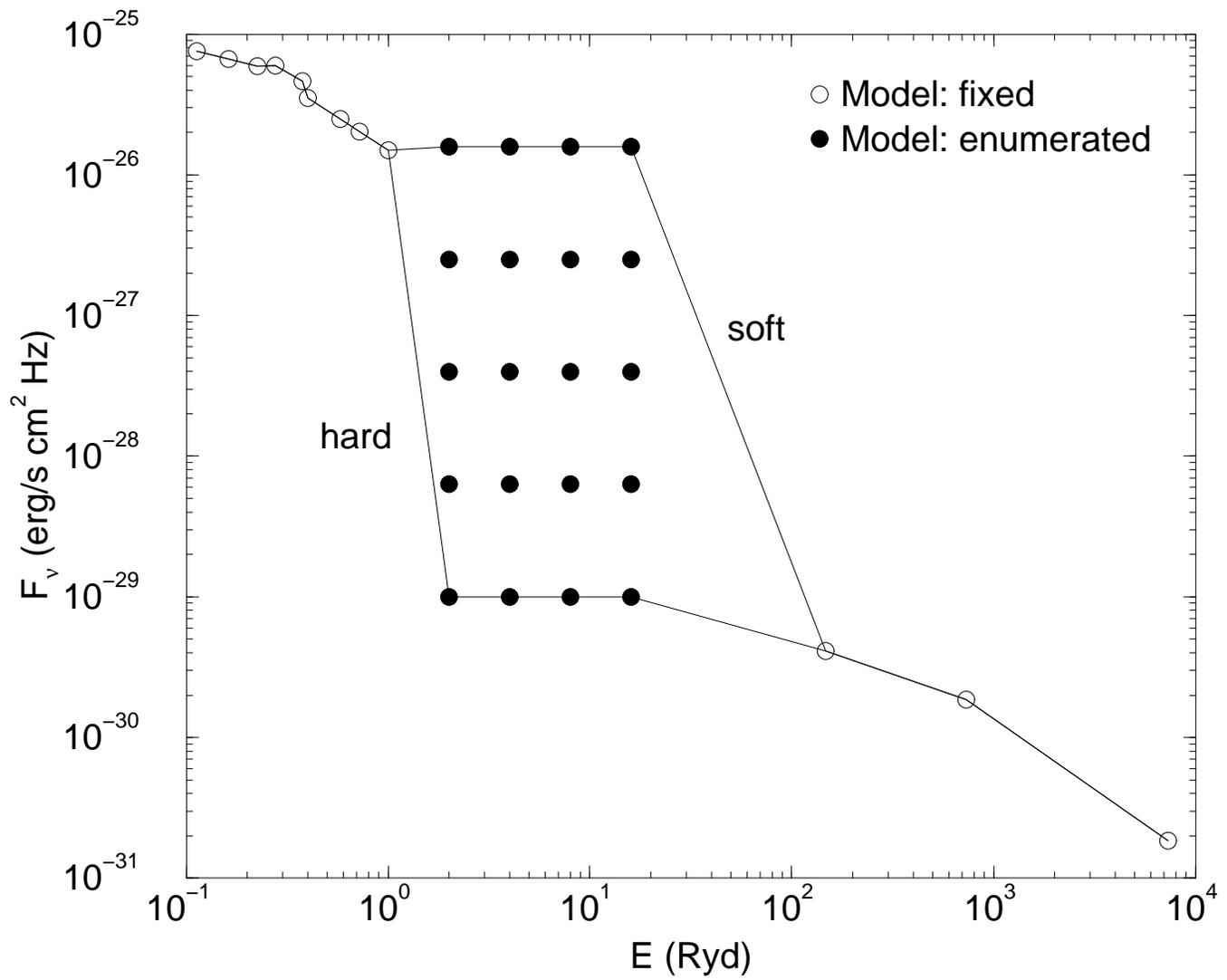
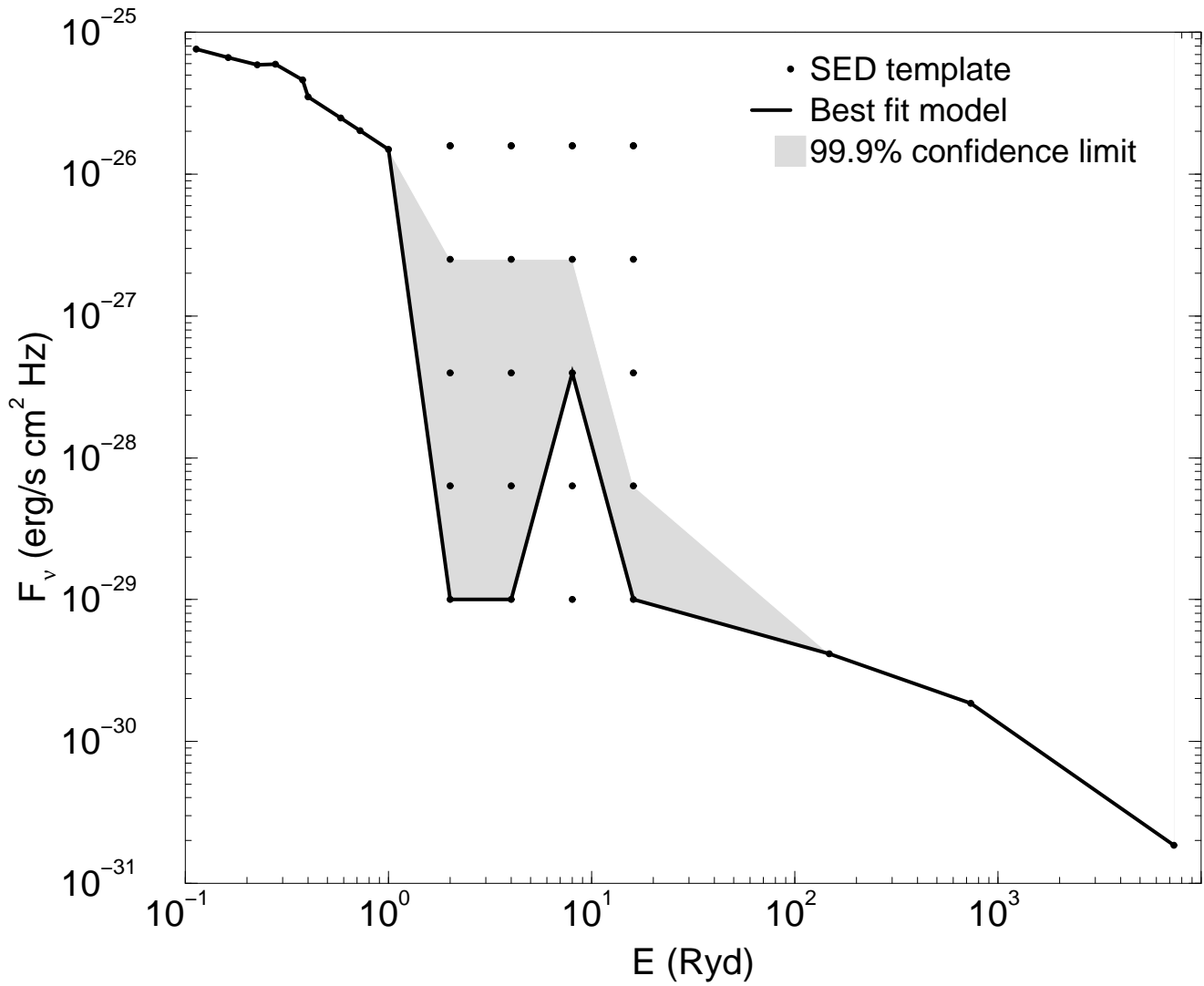
\end{document}